\documentclass[twocolumn,useAMS,usenatbib]{mn2e}

\usepackage{color}
\usepackage{amsmath}
\usepackage{graphicx}
\usepackage{amssymb}
\usepackage{psfrag}

\newcommand{\beq}{\begin{equation}}
\newcommand{\eeq}{\end{equation}}
\newcommand{\bay}{\begin{array}}
\newcommand{\eay}{\end{array}}
\newcommand{\beqa}{\begin{align}}
\newcommand{\eeqa}{\end{align}}
\newcommand{\nn}{\nonumber}
\newcommand{\rmd}{\mathrm{d}}
\newcommand{\brac}[1]{\left({#1}\right)}

\newcommand{\pd}[2]{\frac{\partial{#1}}{\partial{#2}}}
\newcommand{\td}[2]{\frac{\rmd{#1}}{\rmd{#2}}}
\newcommand{\curl}{\nabla\times}
\renewcommand{\div}{\nabla\cdot}

\newcommand{\bB}{{\bf B}}
\newcommand{\bH}{{\bf H}}
\newcommand{\pom}{\varpi}
\newcommand{\be}{{\bf e}}
\newcommand{\br}{{\bf r}}
\newcommand{\fmag}{\boldsymbol{\mathfrak{F}}_{\textrm{mag}}}
\newcommand{\ve}{\varepsilon}
\newcommand{\eps}{\epsilon}
\newcommand{\uj}{\boldsymbol{\hat{\j}}}
\newcommand{\rmp}{\mathrm{p}}
\renewcommand{\rmn}{\mathrm{n}}
\newcommand{\rmx}{\mathrm{x}}
\newcommand{\mup}{\tilde{\mu}_\rmp}
\newcommand{\mun}{\tilde{\mu}_\rmn}
\newcommand{\clE}{\mathcal{E}}

\newcommand{\rhop}{\rho_\rmp}

\newcommand{\skl}[1]{{\color{black}{#1}}}

\title[Magnetic fields in superconducting NSs]
         {The contrasting magnetic fields of superconducting pulsars and magnetars}
\author[S. K. Lander]
       {S. K. Lander${}^1$\thanks{samuel.lander@uni-tuebingen.de}\\ \\
${}^1$ Theoretical Astrophysics, University of T\"ubingen, Auf der
Morgenstelle 10, T\"ubingen 72076, Germany}

\begin{document}

\pagerange{\pageref{firstpage}--\pageref{lastpage}} \pubyear{0000}
\maketitle

\label{firstpage}

\begin{abstract}
We study equilibrium magnetic field configurations
in a neutron star (NS) whose core has type-II superconducting
protons. Unlike the equations for normal matter, which feature no
special field strength, those for superconductors contain the
lower critical field, of order $10^{15}$ G. We find that the ratio
of this critical field to the smooth-averaged stellar field
at the crust-core boundary is the key feature dictating the field
geometry. Our results suggest that pulsar and magnetar-strength fields
have notably different configurations. Field decay for NSs with
$B_\textrm{pole}\sim 10^{14}$ G could thus result in substantial internal
rearrangements, pushing the toroidal field component out of
the core; this may be related to observed magnetar activity. In
addition, we calculate the magnetically-induced ellipticities of our models.
\end{abstract}


\section{Introduction}

The gradual decrease in the rotation rate of a neutron star is
attributed to magnetic dipole radiation and is used to produce an inferred value
for the polar-cap field strength.  This value is often colloquially referred to as `the
star's magnetic field', whereas it is just an estimate of one
component (the dipole) at one location. In the
absence of more detailed information, it is this surface dipole value
--- perhaps with an educated guess about the interior --- which is used to
try to understand magnetic phenomena in neutron stars: the
evolution of their rotation and temperature, the presence or absence
of possible magnetic instabilities
\citep{tayler,wright,akgun_wass,akgun_rmm},
crust-core coupling \citep{easson79} and
precession \citep{mestel_takh}, to name a few. In magnetars, the most highly magnetised objects
known, the magnetic field provides the main reservoir of `free' energy and is
likely to play a major role in their observed flaring activity and
quasi-periodic oscillations \citep{thom_dunc96}.

The problem is that we have no guarantee that the inferred polar-cap
fields of neutron stars are comparable with their interior fields. Two
opposing, but quite feasible scenarios would suggest not. If the
star's magnetic field is mainly confined to the crust the polar-cap
value would be larger than the average interior field; by contrast, if
the field is mostly buried within the star, for example through accretion, the polar-cap
value would be an underestimate. It has been suggested that understanding
these interior fields would help to unify the various observationally
disparate classes of NS \citep{kaspi,vigano}.

To date we have clues about a neutron star's interior field, but
no observations that provide an unambiguous probe of it.  Instead, we
must resort to finding candidate configurations through
modelling. One approach would be an evolutionary one, trying to follow
the star from its birth in a supernova, through any early convective
or dynamo phases \citep{spruit} and the later secular evolution of
the system \citep{pons_mu}. In this work we attack the
problem from a different angle, seeking solutions to a model neutron
star in dynamical equilibrium. To be astrophysically relevant
solutions, these equilibria would also have to be stable, although we
do not study that here.

Specifically, we aim to account for proton superconductivity in a
NS's core, and the effect this has on magnetic-field
equilibria. As discussed in the next section, it has been believed for
decades that NSs contain a large superfluid-superconducting inner region, and magnetic fields are known to
behave very differently in superconducting and normal matter. Despite
this, most studies of NS equilibria are rooted in the classical
physics of normal conductors. Following a brief discussion on
superconductivity we describe accounting for this in our NS model,
\skl{which consists of two effectively decoupled fluids: protons and
  neutrons each obey their own polytropic relation and magnetic forces
  act only on the protons; if we included entrainment there would be
  an effective force on the neutrons too.} We focus
on our treatment of the crust-core boundary, \skl{which is designed to
  minimise any current sheet appearing in this region. Our new
  boundary treatment} provides the main difference in approach between
this work and the preceding paper \citep{sc_eqm_letter}, \skl{where we
  had implicitly assumed magnetar-strength stellar fields, around
  $\sim 10^{15}$ G at the crust-core boundary. After discussing our
  method of solution} we present
results for poloidal and mixed poloidal-toroidal field 
configurations, comparing weak and strong field configurations in a
superconductor with corresponding normal-matter models. \skl{We
  conclude by discussing the possible implications of this work for magnetar activity.}

\section{Superconductivity in neutron stars}
\label{supercon_summary}

There are a number of compelling reasons to believe that the interior of a neutron star
contains a neutron superfluid and a superconducting proton fluid. From
the theoretical side, the broad picture we have of the core region
remains the one outlined by the seminal work of 
\citet{baym_pp}, but the first predictions of superfluidity in neutron
stars \citep{migdal,ginz_kirzh} actually predated the discovery of
pulsars. Although NSs are \skl{certainly} hot by terrestrial standards,
around $10^8$ K, this is in fact extremely cold for such dense
objects; their Fermi temperature is about $10^{12}$ K. NS
matter is therefore highly degenerate, and the microscopic
theory for terrestrial superconductivity \citep{BCS} implies that both
neutrons and protons will form Cooper pairs, leading to a neutron
superfluid and proton superconductor.

Before discussing the details of superconductivity in neutron stars,
we first outline a few key results in the theory of terrestrial
low-temperature superconductivity (two good introductory texts on the
topic are \citet{tinkham} and \citet{tilley2}). There are two important lengthscales in a superconductor: the
penetration depth $\lambda$, the characteristic lengthscale over which
the Meissner effect causes exponential drop-off of the field strength as one moves into the
superconducting region; and the coherence length $\xi$, the
characteristic lengthscale for variation of the
superconducting-particle wavefunction, or the `radius of the Cooper
pair'. The ratio of these two lengthscales forms the Ginzburg-Landau
parameter
\beq
\kappa_{GL}=\frac{\lambda}{\xi}
\eeq
and dictates the type of superconductivity. If
$\kappa_{GL}<1/\sqrt{2}$ type-I superconductivity prevails: the
surface energy of the superconductor is positive and large regions are
devoid of magnetic flux. If instead $\kappa_{GL}>1/\sqrt{2}$ we have a
type-II superconductor: the surface energy is negative and the
superconductor is unstable to the increase of its surface area
through the formation of fluxtubes, provided that the magnetic field
strength $B>H_{c1}$,
the `lower critical field'. The magnetic field is thus able to
penetrate the bulk of the superconductor through these fluxtubes. As
the field strength is increased the 
fluxtubes become more densely packed, until finally at $B=H_{c2}$ (the
`upper critical field') superconductivity is destroyed.

Returning to astrophysics, the longest-standing piece of evidence in favour of superfluidity in
neutron star cores is pulsar glitches --- rapid events in which the
star is seen to spin up slightly (in contrast to its long-term
decrease in rotation rate). This is widely thought to represent a
transfer of angular momentum from a more rapidly-spinning internal superfluid
component to the crust \citep{and_itoh}. More recently, the rapid
cooling of the Cassiopeia A neutron star has been attributed to the
onset of neutron superfluidity, together with protons already in a
superconducting state \citep{page_letter,shternin_letter}. Even at magnetar field strengths, neutron stars are believed to become cold
enough to admit superfluid-superconducting regions within a few
hundred years of their birth \citep{ho_glam_and}.

The protons of a NS's outer core are
expected to form a type-II superconductor, with values for the two critical fields
being roughly $H_{c1}\sim 10^{15}$ G and $H_{c2}\sim
10^{16}$ G, but the timescale for Meissner expulsion is so long that
the fluxtube state is still expected to be metastable when $B<H_{c1}$
\citep{baym_pp,jones87}. Because of its regularity, one can take an
average over this fluxtube array to determine the macroscopic magnetic force acting on a
type-II superconductor \citep{easson_peth,mendell91,akgun_wass,GAS}:
\beq \label{fmag_supercon}
\fmag = -\frac{1}{4\pi}\Bigg[
                  \bB\times(\curl{\bf H}_{c1})
                   +\rhop\nabla\brac{B\pd{H_{c1}}{\rho_\rmp}}
                                  \Bigg],
\eeq
where $\bB$ represents the smooth-averaged magnetic field and ${\bf
  H}_{c1}=H_{c1}\hat\bB$ ($\hat\bB=\bB/B$, the magnetic field unit
vector). This force is clearly not simply a rescaling of the
normal-matter Lorentz force. There is also likely to be an inner core region
with type-I superconducting protons, with alternating $B=0$ and $B\neq 0$ domains, but
without the regular structure of a fluxtube array
\citep{sedrakian}. As yet there is no equivalent averaged magnetic 
force for this case.

\section{Neutron star model}

\subsection{Assumptions}

We model a neutron star as a self-gravitating, stationary and
axisymmetric fluid body, with parallel rotation and magnetic symmetry
axes. We will adopt cylindrical polar coordinates $(\varpi,\phi,z)$,
aligning the star's symmetry axis with the coordinate $z$-axis. The
star is non-relativistic and has zero electrical resistivity. It has a
core composed of type-II superconducting protons, superfluid
neutrons and normal electrons; this is effectively a two-fluid system, 
since the inertia of the electrons is negligible and their chemical
potential can be absorbed into that of the protons. Beyond a certain
density in a neutron star, this admixture of particles may give way to
an inner core with exotic particles or a region with type-I
superconductivity. We ignore any such physics for simplicity, but it
is possible that such high densities may not be attained in large
regions of the star anyway; the recently observed two-solar-mass pulsars 
suggests neutron stars obey a fairly stiff equation of state with
relatively low central density \citep{demorest,antoniadis}.

The core is surrounded by an elastic crust of normal
matter, which we assume to be relaxed. Note that a stressed crust
evolves secularly in a manner governed by Hall drift and Ohmic decay ---
with the former term expected to be dominant, the crustal field would
eventually settle into a `Hall equilibrium' configuration
\citep{gourgouliatos}. Since we assume there are no elastic stresses,
the equations for the crust reduce to those of a single fluid. We can,
therefore, model the entire star as a two-fluid equilibrium:\\
1. the superfluid neutrons extend from the centre of the star to the
crust-core boundary;\\
2. the `proton' fluid represents the proton-and-electron fluid in the
core, where it is subject to the fluxtube tension force;\\
3. beyond the crust-core boundary, the same `proton' fluid is
reassigned to become the single-fluid crust, representing an
unstrained ion lattice and obeying the Lorentz force of
normally-conducting matter.

Our two-fluid model gives us the freedom to choose different pressure-density
relations for the neutrons and protons, making us better able to mimic
a `realistic' NS profile than with a polytropic
equation of state \citep{LAG}. It also enables us to introduce
composition-gradient stratification in the stellar core. For very young
neutron stars it would be more appropriate to allow for stratification
from temperature/entropy gradients
\citep{reis_strat,mastrano,yoshida}; in this case the star would also 
be too hot to form the multifluid state studied here. Our work is
applicable to neutron stars over a few hundred years old, 
including magnetars \citep{ho_glam_and}.

\subsection{General equations}
\label{general_eqs}

As described above, our neutron star model is a two-fluid system in
equilibrium. This is therefore governed by a pair of Euler equations
--- one for the neutrons (denoted with a subscript n) and another for
the `proton' fluid (subscript p), used to describe the combined
proton-electron fluid in the core and the relaxed ion lattice in the crust:
\beq \label{n_Euler}
\nabla\brac{\mun+\Phi-\frac{\pom^2\Omega_\rmn^2}{2}}=0,
\eeq
\beq \label{p_Euler}
\nabla\brac{\mup+\Phi-\frac{\pom^2\Omega_\rmp^2}{2}}=\frac{\fmag}{\rho_\rmp},
\eeq
where $\tilde\mu_\rmx$ is the chemical potential of each particle species
($\rmx\in\{\rmn,\rmp\}$), $\Phi$ gravitational potential, $\Omega_\rmx$
each fluid's rotation rate and $\fmag$ the magnetic force; the form of
this force will be the only difference between the normal and
superconducting cases. We are also implicitly assuming there is no
entrainment, a coupling between particle species which would lead to a
magnetic force on the neutrons for a superfluid-superconducting
core \citep{GAS}. We will only consider the case of corotating fluids, 
i.e. $\Omega_\rmn=\Omega_\rmp\equiv\Omega$.

Even in the absence of any
other coupling mechanisms, the two fluids are indirectly linked
through Poisson's equation, as they both feel the same gravitational potential:
\beq \label{Poisson}
\nabla^2\Phi = 4\pi G\rho = 4\pi G(\rho_\rmn+\rho_\rmp).
\eeq
Instead of using the two individual Euler equations, we will work with
the neutron-fluid Euler and a `difference-Euler', given
by \eqref{p_Euler} minus \eqref{n_Euler}:
\beq \label{d_Euler}
\nabla\brac{\mup-\mun} = \frac{\fmag}{\rho_\rmp}.
\eeq
Taking the curl of this equation shows that $\fmag/\rhop$ is
irrotational and so (as in the single-fluid case) there exists a
scalar $M$ such that
\beq \label{magforce_M}
\frac{\fmag}{\rho_\rmp}=\nabla M.
\eeq
The magnetic force --- whether the normal-matter Lorentz force or the
superconducting fluxtube tension --- therefore depends on a scalar function
$M$ and the proton-fluid density $\rhop$, as opposed to the single-fluid case
where the dependence is on the total mass density: $\fmag/\rho=\nabla
M$. The single-fluid versions of the magnetic equations we present here may
be found simply by replacing any $\rhop$ terms with the total $\rho$.

Another universal result is the Maxwell equation
\beq
\div\bB=0.
\eeq
Since our system is axisymmetric we may use this equation to
write $\bB$ in terms of a streamfunction $u$:
\beq \label{B-u}
\bB = \frac{1}{\pom}\nabla u\times\be_\phi + B_\phi\be_\phi.
\eeq
This implies that whatever the magnetic force, we have $\bB\cdot\nabla
u=0$; field lines are parallel to contours of the streamfunction.

We close our system with an equation of state.  As in previous papers
\citep{LAG,sc_eqm_letter}, we choose a two-fluid generalisation of a polytrope. Specifically, we
choose an energy functional
\beq \label{EOS}
\clE = \clE(\rho_\rmn,\rho_\rmp,w_{\rmn\rmp}^2)
       = k_\rmn\rho_\rmn^{1+1/N_\rmn}
           + k_\rmp\rho_\rmp^{1+1/N_\rmp}.
\eeq
In general the functional would involve a term in the relative velocity
between the particle species
${\bf w}_{\rmn\rmp}\equiv{\bf v}_\rmn-{\bf v}_\rmp$, coupling the two
fluids (like entrainment); in our simple choice for $\clE$ this term
is zero. The chemical potentials of the species are now defined
through $\mathcal{E}$:
\beq
\tilde\mu_{\textrm{x}}\equiv\left.\pd{\clE}{\rho_{\textrm{x}}}\right|_{\rho_{\textrm{y}}},
\eeq
where $\rmx$ represents one particle species and ${\textrm{y}}$ the other.

For our equation of state, then, each particle species obeys its own
barotropic relation: $\tilde\mu_\rmn=\tilde\mu_\rmn(\rho_\rmn)$ and
$\tilde\mu_\rmp=\tilde\mu_\rmp(\rho_\rmp)$. Our chosen energy functional
may therefore be thought of as a `double polytrope' in the two fluids
\citep{prix_ca}. \skl{Note that the total pressure is not generally
  a function of the total density though, $P\neq P(\rho)$.} Our
equation of state enables us to derive results for a stratified
two-fluid star which are equivalent to those for a single-fluid
barotropic star, although one should be aware that the introduction of
proton-neutron coupling could lead to qualitative changes in the
resulting equilibria.

\subsection{Normal crust}

As described earlier, we are treating the neutron star crust as being
relaxed, so that its equilibrium state involves only fluid and
magnetic forces, with no shear stresses. Unlike the core protons, the
crust is expected to be composed of normally-conducting matter, which
we treat as a single fluid. In this familiar case $\fmag$ is the Lorentz force, given by
\beq
\fmag=\frac{1}{4\pi}(\curl\bB)\times\bB.
\eeq
For axisymmetric equilibria, the behaviour of the magnetic field may
be encapsulated in a single differential equation, the
\emph{Grad-Shafranov equation} \citep{grad_rubin,shafranov}:
\beq \label{grad_shaf}
\Delta_*u = -4\pi\rhop\varpi^2\td{M}{u} - f_N\td{f_N}{u},
\eeq
where $f_N$ is related to the toroidal-field component through
\beq \label{f_normal}
f_N(u)=\varpi B_\phi,
\eeq
and $\Delta_*$ is a differential operator:
\beq \label{operator}
\Delta_*
 \equiv \pd{^2}{\varpi^2}-\frac{1}{\varpi}\pd{ }{\varpi}+\pd{^2}{z^2}
\eeq
--- this is the Laplacian operator with the sign of the
second term reversed. For later comparison, we note from equations
\eqref{B-u} and \eqref{f_normal} that the magnitude of the magnetic field is given by
\beq \label{B_normal}
B = \frac{\sqrt{|\nabla u|^2+f_N^2(u)}}{\varpi}.
\eeq
The Grad-Shafranov equation
describes mixed poloidal-toroidal fields, or purely
poloidal fields by taking $f_N(u)=0$. In the special case of a purely
toroidal field there is no separate equation, and the Lorentz force
term in the Euler equation reduces to the form:
\beq
M=-\frac{1}{4\pi}\int_0^{\rhop\varpi^2}
           \frac{\eta_N(\beta)}{\beta}\td{\eta_N}{\beta}\ \rmd\beta,
\eeq
for some free function $\eta_N=\eta_N(\rhop\varpi^2)$. This function
is related to the magnetic field strength through
\beq
B_\phi=\frac{\eta_N(\rhop\varpi^2)}{\varpi}.
\eeq
A detailed derivation of these relations is given in \citet{LJ09}.

\subsection{Type-II superconducting core}

We assume that the protons throughout the core form a
type-II superconductor, with the magnetic field quantised into
fluxtubes, resulting in the magnetic force \eqref{fmag_supercon} discussed in
section \ref{supercon_summary}. Although this force is more complicated than the Lorentz
force, we may use many of the same tricks as for the Grad-Shafranov
equation. Again, the problem splits into two cases: one for purely
toroidal fields and another for poloidal/mixed fields. We relegate the
full derivations to appendix \ref{supercon_GS_deriv} and report just the key results
here.

\subsubsection{Toroidal fields}

The case of purely toroidal fields in a superconductor is very similar
to that for normal matter. There is no separate equation for the
magnetic field; the proton Euler equation simply gains one extra term
related to the magnetic force (recall that $\nabla M=\fmag/\rhop$):
\beq
M = -\frac{h_c}{4\pi}\int_0^{\rhop\varpi}
                \frac{1}{\beta}\td{\eta}{\beta}\ \rmd\beta
\eeq
for some free function $\eta=\eta(\rhop\varpi)$. The corresponding
magnetic field strength is
\beq
B_\phi=\frac{\eta(\rhop\varpi)}{\rhop\varpi}.
\eeq
Configurations with this type of magnetic field were studied in
earlier work \citep{akgun_wass,LAG} and are not the subject of the present study; we include
the relations for the sake of completeness.

\subsubsection{Mixed poloidal-toroidal fields}

The derivation of an equation governing mixed poloidal-toroidal fields
in a superconductor is similar, in many ways, to that for the
normal-matter Grad-Shafranov equation. One key difference
is that the magnetic-force function $M$ (see equation
\eqref{magforce_M}) is no longer a function of the streamfunction 
$u$. Since our desired end result is an equation in terms of $u$ and
functions thereof, we instead define a related quantity which
involves $M$ but is also a function of $u$:
\beq \label{y_defn}
y(u) = \frac{4\pi M}{h_c}+B.
\eeq
The local field strength $B$ appears many
times, in combination with the proton density, so for brevity we define the
quantity
\beq
\Pi\equiv\frac{B}{\rhop}.
\eeq
Using these quantities, we show in appendix \ref{supercon_GS_deriv} that a mixed
poloidal-toroidal field is governed by the following equation:
\beq \label{supercon_GS}
\Delta_* u -\frac{\nabla\Pi\cdot\nabla u}{\Pi}
  = -\varpi^2\rhop\Pi\td{y}{u} - \Pi^2 f\td{f}{u}.
\eeq
The free function $f(u)$ is again related to the toroidal field
component, but differs from the normal-matter result by a factor of
$\Pi$:
\beq \label{f_supercon}
f(u)=\frac{\varpi B_\phi}{\Pi}.
\eeq
Equation \eqref{supercon_GS} is the equivalent of the
Grad-Shafranov equation when the stellar matter (protons in our
two-fluid system) is a type-II 
superconductor. The most significant difference from the normal case
is the presence of terms related to the field strength $B$ (in the
quantity $\Pi$). Using equations \eqref{B-u} and \eqref{f_supercon}, we see that
$B$ is indirectly related to $u$:
\beq \label{B_supercon}
B=\frac{|\nabla u|}{\sqrt{\varpi^2-f^2(u)/\rhop}},
\eeq
which may be compared with the normal-matter result
\eqref{B_normal}. The presence of $B$ therefore represents an
additional degree of nonlinearity in the equation; in attempting to
solve \eqref{supercon_GS}, we may
anticipate particular difficulties associated with the
$\nabla\Pi\cdot\nabla u/\Pi$ term. Another difference is that whereas the 
magnetic force appears explicitly in the Grad-Shafranov equation,
through the function $M$, it does not in the corresponding equation
for superconducting matter.

The results reported in this section agree with expressions in \citet{hen_wass}, who
specialised to poloidal fields and used a
different method of solution, which did not require the
explicit form of equation \eqref{supercon_GS} in terms of the
streamfunction.

\subsection{Crust-core boundary conditions}

We need to solve for the magnetic field in three distinct domains: the
superconducting core, normal crust and vacuum outside the star. In a
spherical model we would be able to put the surface and crust-core boundaries at
some specified radius, but since we wish to allow for deformations of
the star we define these boundaries
in terms of proton density contours instead. Denoting the equatorial surface
by $r_{\textrm{eq}}$, the stellar surface and crust-core boundary are given by
the pairs of $(\varpi,z)$ coordinates satisfying
\begin{align}
\rhop^{\textrm{surface}}(\varpi,z) &= \rhop(r_{\textrm{eq}},0) = 0, \\
\rhop^\textrm{cc}(\varpi,z)                   &= \rhop(0.9r_{\textrm{eq}},0).
\end{align}
The magnetic equation to be solved is then:
\beq \label{sc_normal_ext_regions}
\Delta_* u = \begin{cases}
                      \displaystyle
                       -\varpi^2\rhop\Pi y'(u) - \Pi^2 f(u)f'(u) 
                            + \frac{\nabla\Pi\cdot\nabla u}{\Pi}& 
                          \textrm{core}\\
                       -4\pi\varpi^2\rhop M_N'(u) - f_N(u)f_N'(u)  & 
                          \textrm{crust}\\
                       0  & \textrm{exterior}
                     \end{cases}
\eeq

At the crust-core boundary of a neutron star there is likely to be complicated
microphysics, with forces pinning fluxtubes to the crust and perhaps a
current sheet. Since we do not yet have a good model for these
effects, we consider it safer first to omit these and model the simplest crust-core
boundary: requiring force balance, and for the field to remain divergence-free (equivalent to imposing continuity of the
field component perpendicular to the boundary). The latter condition is satisfied
automatically by working with the streamfunction. \skl{A more detailed
  discussion of the crust-core boundary, including aspects of the
  microphysics of the region, is given in \citet{hen_wass}.}

\subsubsection{Earlier treatment of boundary}
\label{BC_old}

In \citet{sc_eqm_letter}, force balance was ensured
by arguing that whilst $\bB$ and $\bH$ are distinct quantities in
the superconducting core, they should be equal in the crust. In the
core
\beq
\bH_{c1}=\frac{H_{c1}}{B}\bB;
\eeq
this can be matching smoothly to the crust by requiring that
$\bH_{c1}\to\bH^\textrm{crust}=\bB^\textrm{crust}$ at the boundary,
i.e. $H_{c1}\to B$. This was done by making appropriate choices of the
core functions $y(u),f(u)$ in terms of the crust functions
$M_N(u),f_N(u)$. This only allows for smooth matching
without an associated surface current provided that $B^\textrm{cc}\sim
10^{15}$ G, however; potentially appropriate only for
magnetars. \skl{Assuming $H_{c1}\to B$ at the crust-core boundary
  effectively prevented us from finding solutions with a sharp
  transition, like those of \citet{hen_wass}.}

\subsubsection{New treatment of boundary}
\label{BC_new}

\skl{In this paper we use a more general treatment than that of
  \citet{sc_eqm_letter}, allowing us to study the
more typical case where $B<H_{c1}$ at the boundary. Previous work on
poloidal fields in superconducting stars specialised to this case, using geometric techniques to
solve along individual field lines and formulating the boundary
conditions accordingly: \citet{roberts} argued that to
match a superconducting region to vacuum, all field lines should cross
the surface perpendicularly, whilst \citet{hen_wass} discuss a class
of solutions where field lines hit the boundary of the superconducting
domain vertically. Our approach is different and does not allow us to
specify the geometry of field lines \emph{a priori}.}

At the boundary we want the magnetic force in the core and the Lorentz
force in the crust to balance. Using square brackets and a
superscript `cc' to denote evaluation at the crust-core boundary, we have:
\beq \label{force_bal}
\left[\rhop^\textrm{core}\nabla M_{sc}\right]^\textrm{cc} = \left[\rhop^\textrm{crust}\nabla M_N(u)\right]^\textrm{cc}.
\eeq
At this stage we allow for a possible jump in the density at the crust-core
boundary (expected from more realistic equations of state, e.g. \citet{akmal_pr}), though in our models $\rhop$
is smooth everywhere. Using the definition of $y(u)$, we may rewrite this as
\beq
\left[\nabla B^\textrm{core}\right]^\textrm{cc} = 
 \left[ \brac{\td{y}{u}-\frac{4\pi}{h_c}\frac{\rhop^\textrm{crust}}{\rhop^\textrm{core}}\td{M_N}{u}}\nabla u \right]^\textrm{cc}.
\eeq
Dotting $\bB$ into both sides, we see that $B^\textrm{core}$ at the crust-core
boundary should be a function of the
streamfunction $u$, since $[\bB\cdot\nabla B^\textrm{core}]^\textrm{cc}=\bB\cdot\nabla u=0$.
This is not straightforward to enforce, because the crust-core boundary is given by a
contour of $\rhop$, not of $u$. We will look for a polynomial
approximation $\tilde{B}^\textrm{cc}(u)$ to the actual value of $B$ along the
core side of the boundary, as evaluated using relation \eqref{B_supercon},
which we term $B^\textrm{cc}$.
Let us assume a quadratic relation between $\tilde{B}^\textrm{cc}(u)$ and $u$:
\beq
\tilde{B}^\textrm{cc}(u)=c_0+c_1u+c_2u(u-u^\textrm{cc}_{\textrm{eq}}),
\eeq
where $c_0,c_1,c_2$ are constants and $u^\textrm{cc}_{\textrm{eq}}$ is the value of
$u$ at the equatorial crust-core boundary. Now, the streamfunction $u$
attains its maximum in the centre of the 
closed-field line region and is always zero along the pole --- at this
latter location we have $\tilde{B}^\textrm{cc}(u)=c_0$ and so we choose
\beq
c_0=B^\textrm{cc}_\textrm{pole}.
\eeq
We now ensure agreement between $\tilde{B}^\textrm{cc}(u)$ and $B^\textrm{cc}$ at the
equator by choosing
\beq
c_1=\frac{B^\textrm{cc}_{\textrm{eq}}-c_0}{u^\textrm{cc}_{\textrm{eq}}}.
\eeq
Finally, within our quadratic approximation we can also ensure that
$\tilde{B}^\textrm{cc}(u)$ and $B^\textrm{cc}$ coincide at some middle point (e.g. at
$\theta=\pi/4$) by fixing $c_2$:
\beq
c_2=\frac{B^\textrm{cc}_\textrm{mid}-c_0-c_1u^\textrm{cc}_\textrm{mid}}{u^\textrm{cc}_\textrm{mid}(u^\textrm{cc}_\textrm{mid}-u^\textrm{cc}_{\textrm{eq}})}.
\eeq

\begin{figure}
\begin{center}
\begin{minipage}[c]{1.0\linewidth}
\psfrag{magnitude}{magnitude (dimensionless units)}
\psfrag{theta}{$\theta$}
\psfrag{u}{$u$}
\psfrag{B_actual}{$B^\textrm{cc}$}
\psfrag{B_u}{$\tilde{B}^\textrm{cc}(u)$}
\includegraphics[width=\linewidth]{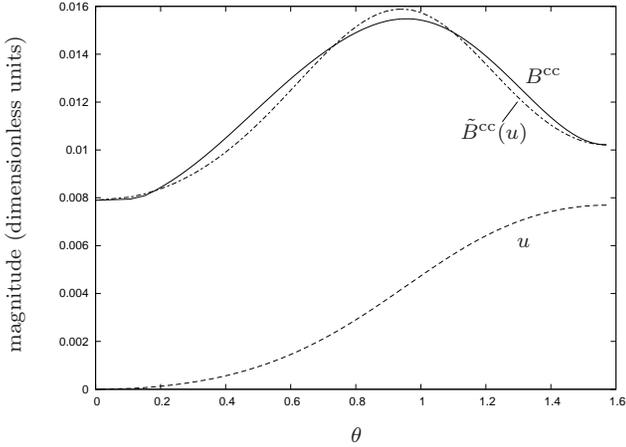}
\end{minipage}
\caption{\label{Bcc_matching}
               Quantities along the crust-core boundary, from the pole
               ($\theta=0$) to the equator ($\theta=\pi/2$). To impose
               force balance across the boundary we need to find a
               function $\tilde{B}^\textrm{cc}(u)$ of the streamfunction $u$
               which approximates the actual value $B^\textrm{cc}$ of the
               magnetic field there; see section \ref{BC_new}.}
\end{center}
\end{figure}

Over successive steps within our iterative scheme (see section
\ref{appendix_numerics} for details) we have found this expression for $\tilde{B}^\textrm{cc}(u)$
provides a moderately good approximation to $B$ along the
crust-core boundary, with a discrepancy typically less than $10\%$;
see figure \ref{Bcc_matching}. Any discrepancy would result in an
undesired, unmodelled current sheet at the boundary, so we wish to minimise this.
We are now able to ensure force balance at the crust-core boundary by
choosing
\beq \label{y-from-MN}
y(u)=\tilde{B}^\textrm{cc}(u)+\frac{4\pi}{h_c}\left[\frac{\rhop^\textrm{core}}{\rhop^\textrm{crust}}\right]^\textrm{cc}M_N(u).
\eeq

Next, for a mixed poloidal-toroidal field the crust-core discontinuity in
$B_\phi$ is given by: 
\beq
\left[B^\textrm{core}_\phi-B^\textrm{crust}_\phi\right]^\textrm{cc}
   =\left[\frac{1}{\varpi}\brac{\frac{B^\textrm{core}}{\rhop^\textrm{core}}f_{sc}(u)-f_N(u)}\right]^\textrm{cc}.
\eeq
and so we can ensure continuity of the toroidal component by choosing
\beq \label{fsc-from-fN}
f(u)=[\rhop^\textrm{core}]^\textrm{cc}\frac{f_N(u)}{\tilde{B}^\textrm{cc}(u)}.
\eeq

Our original system of equations \eqref{sc_normal_ext_regions}
contained four free functions of the streamfunction. To avoid
any current sheets at the boundary, however, equations \eqref{y-from-MN} and
\eqref{fsc-from-fN} tell us that only two of these may be chosen
independently. We will choose the functional forms in the crust $M_N(u),f_N(u)$
and use these to fix the core functions $y(u),f(u)$; mathematically,
it would be equivalent to choose the core functions and fix those in
the crust, but this proved to be less numerically stable.

\subsection{Numerics}

We cast the magnetic-field equation \eqref{sc_normal_ext_regions} and
the other equilibrium equations \eqref{n_Euler}, \eqref{Poisson} and \eqref{d_Euler} into
integral form and solve them using a non-linear iterative scheme. The
foundation of our code is the self-consistent field method
\citep{hachisu}, but extended considerably to solve for superfluid-superconducting stars with
magnetic fields. Since our method is not perturbative, we can
consistently allow for the back-reaction of the field on the fluid. We
are also able to include the contributions from higher multipoles with
no additional difficulty; we typically perform a sum up to an angular
index $l=16$. A more detailed description is given in appendix
\ref{appendix_numerics}.

Within the code we use dimensionless variables, non-dimensionalising
all quantities by using an appropriate combination of powers of the
gravitational constant, the equatorial surface radius and
the total central density. For all results with
physical units, we have redimensionalised to a $1.4$-solar-mass NS
with a radius of $10$ km; more details are given in \citet{LAG}.

\section{Results}

\subsection{Choices for the free functions}

Within our model of a superconducting neutron star, the magnetic field
is specified by the single equation \eqref{sc_normal_ext_regions}. As
discussed above, crust-core boundary conditions mean that just two of the four
functions of the streamfunction may be chosen independently, with the
other two related to these. These free functions are $M_N(u)$,
dictating the strength and distribution of the magnetic force
throughout the star (indirectly related to the poloidal field
component), and $f_N(u)$, dictating the strength of the toroidal-field
component.

Some freedom is possible in choosing these, although their
effect on the final equilibrium is not dramatic. Since equation
\eqref{sc_normal_ext_regions} involves the derivative $M'_N(u)$, we
need $M_N(u)$ to be a power of $u$ greater than or equal to unity;
indeed, we will usually take $M_N(u)=\kappa u$, where $\kappa$ is a
constant which will set the magnitude of the magnetic force (and hence
the field strength). The simplest choice for the other function is
$f_N(u)=0$, which leads to a purely poloidal field. For mixed-field
configurations, we need to ensure $f_N(u)$ is zero outside the star,
to avoid an exterior toroidal field and corresponding electric
current\footnote{Real neutron stars \emph{do} have an exterior
  current distribution, the magnetosphere, and $f_N(u)$ could be
  chosen to account for this \citep{GLA}. Here we focus for simplicity
  on the interior equilibrium and assume a vacuum exterior.}. This is
done by fitting $f_N(u)$ to $u_\textrm{int}$, the largest $u$-contour (i.e. field line)
which closes within the star:
\beq \label{fN_choice}
f_N(u)=\begin{cases}
              a(u-u_\textrm{int})^{\zeta+1}   & {\textrm{for }} u\geq u_\textrm{int}\\
              0                                   &  {\textrm{for }} u< u_\textrm{int}.
            \end{cases}
\eeq
Note that the exponent must, again, be greater than unity, since our
equations also involve $f'_N(u)$. $\zeta$ fixes the steepness of the
toroidal-field profile and is set to $0.1$ in this work; $a$ dictates
the strength of the toroidal component.

\subsection{Purely poloidal fields}

\begin{figure}
\begin{center}
\includegraphics[width=0.8\linewidth]{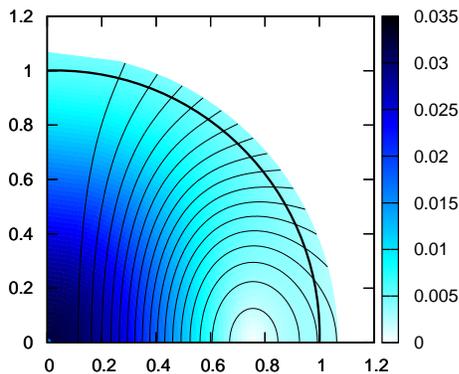}
\caption{\label{normal_pol}
               A poloidal magnetic field in a neutron star with a
               two-fluid core and single-fluid crust, but obeying
               normal MHD in both domains. As for all other
               poloidal-field models, we show the field magnitude
               with the colour scale and its direction with the field
               lines. The stellar surface is shown with the arc at
               dimensionless radius $r=1.0$. The crust-core boundary is at
               $r=0.9$, but to avoid cluttering the figure it is not plotted. The field 
               strength volume-averaged over the star, $\bar{B}$, is $10^{16}$ G here, but the corresponding
               plots for $\bar{B}=10^{14}$ G and $10^{17}$ G are
               indistinguishable.}
\end{center}
\end{figure}

We begin this section with a reminder of a typical poloidal-field
configuration in a \emph{normally-conducting} two-fluid star, in
order to be able to make a direct comparison with our new results for
a superconducting model. In figure \ref{normal_pol} we show the
magnitude and direction of an equilibrium poloidal field in a model NS
subject to the normal-matter Lorentz force in both
core and crust, with $M_N(u)=\kappa u$. The region of field lines that close within the star
is centred at a dimensionless equatorial radius $r\approx 0.75$ and
the transition between core and crust is smooth. The field geometry is
essentially independent of the field strength.

\skl{Note that although all our models involve a two-fluid core, the
  neutrons play a virtually negligible role in the field
  configurations presented in this paper, because their influence on
  the protons is only indirect, through Poisson's equation
  \eqref{Poisson}. The inclusion of entrainment would provide a more
  direct coupling between the two fluids, and we anticipate the
  distribution of the neutrons would then have some direct effect on the
  magnetic field.}

\begin{figure*}
\begin{center}
\begin{minipage}[c]{1.0\linewidth}
\includegraphics[width=\linewidth]{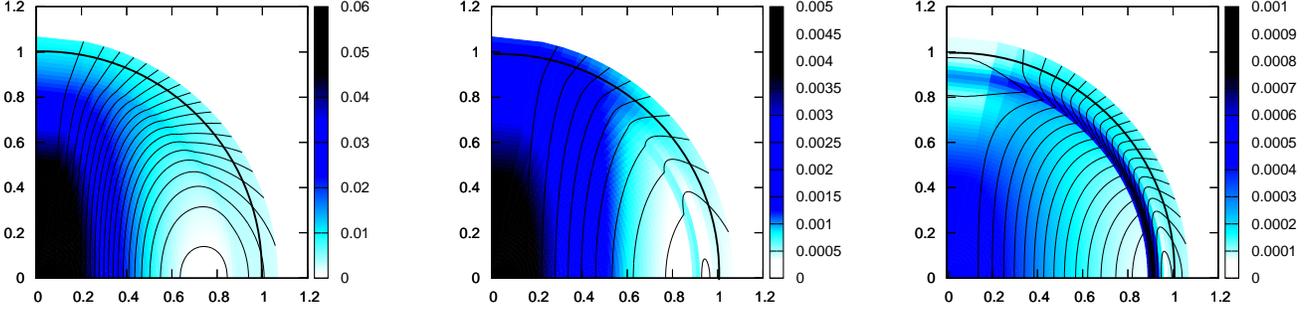}
\end{minipage}
\caption{\label{pol_Mu}
               Poloidal magnetic fields in a star with a
               superconducting core and normal crust, with the
               boundary between the two located at a dimensionless
               radius $r=0.9$ and the surface at $r=1.0$. We choose
               $M_N(u)=\kappa u$. Unlike the normal-matter case in
               figure \ref{normal_pol}, the equilibria here are qualitatively
               different for different field strengths, depending on
               the average ratio of $B$ to $H_{c1}$ at the crust-core
               boundary. The value of this quantity $\langle
               B^\textrm{cc}\rangle/H^\textrm{cc}_{c1}=5.7,0.7,0.4$ for the left,
               middle and right panels, respectively. As this ratio is decreased the
               field lines become sharper across the boundary and the
               closed-field region is pushed out into the crust.}
\end{center}
\end{figure*}

Using the same functional form $M_N(u)=\kappa u$, we now consider a
star whose core protons form a type-II superconductor and hence are
subject to a fluxtube tension force instead of the Lorentz force; the results
are plotted in figure \ref{pol_Mu}. The three models only differ in
their field strength --- in particular, the important criterion is the relative
magnitude of the smooth-averaged field $B$ to the critical field
$H_{c1}$ at the crust-core boundary. $H_{c1}$ is density-dependent and
therefore constant along the boundary, but $B$ is not; for this reason
we use the average value $\langle B^\textrm{cc}\rangle$ for comparisons.

In the left-hand plot of figure \ref{pol_Mu} we show a star where $\langle B^\textrm{cc}\rangle/H^\textrm{cc}_{c1}>1$; in this case the ratio
is $5.7$. The field direction and magnitude are both smooth across the
boundary and the closed-field line region penetrates the core; the
configuration is qualitatively similar to the normal-matter model
shown in figure \ref{normal_pol}. It also resembles the poloidal-field
solution presented in \citet{sc_eqm_letter}, where a strong magnetic
field had been assumed \skl{(see section \ref{BC_old}).}

The middle plot shows a star with 
$\langle B^\textrm{cc}\rangle/H^\textrm{cc}_{c1}\sim 1$ (specifically, $0.7$ in
this case). The field lines kink slightly across the crust-core
boundary and the closed field lines are pushed outwards. Finally, the
right-hand plot is for a star with $\langle B^\textrm{cc}\rangle/H^\textrm{cc}_{c1}<1$
(the ratio is $0.4$ in this particular model), probably the typical
case for pulsars. The closed-field line region has been
completely pushed out into the normally conducting crust, whilst no
field lines close in the core. The field strength is highest in the vicinity of the
crust-core boundary.

\skl{The left-hand plot of figure \ref{pol_Mu}, like figure
  \ref{normal_pol}, shows field lines hitting the crust-core boundary
  perpendicularly in the region of the pole and tangentially at the
  equator. By contrast, in the right-hand plot of figure \ref{pol_Mu}
  all field lines hit the boundary approximately perpendicularly,
  consistent with the expectation of \citet{roberts}. Furthermore,
  with no field lines closing in the core, this} field configuration is
encouragingly similar to both those presented in \citet{hen_wass} and
\citet{roberts}.

\begin{figure*}
\begin{center}
\begin{minipage}[c]{1.0\linewidth}
\includegraphics[width=\linewidth]{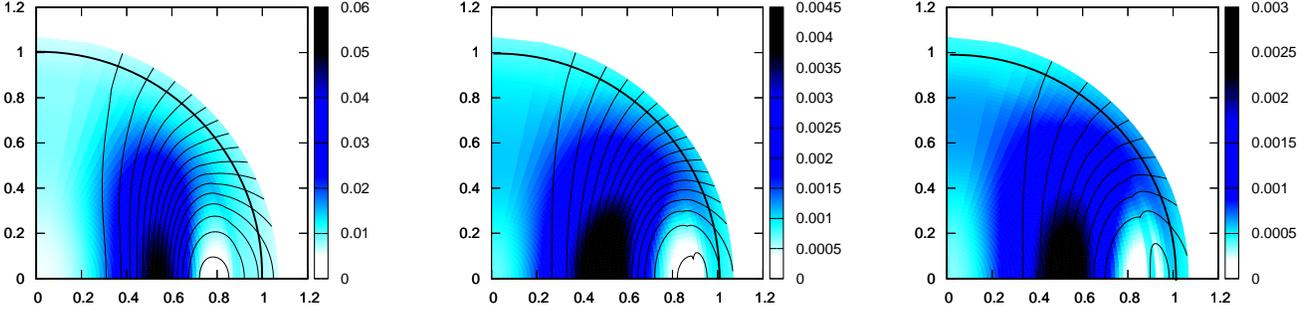}
\end{minipage}
\caption{\label{pol_Muu}
               To check our results from figure \ref{pol_Mu} are not
               specific to the choice of flux function used there, we
               plot the corresponding sequence of equilibria for the case
               $M_N(u)=\kappa u^2$ here. The left, middle and right
               panels represent the cases $\langle B^\textrm{cc}\rangle>H^\textrm{cc}_{c1}$, 
               $\langle B^\textrm{cc}\rangle\sim H^\textrm{cc}_{c1}$ and 
               $\langle B^\textrm{cc}\rangle<H^\textrm{cc}_{c1}$,
               respectively. Although the core field distributions are
               different from figure \ref{pol_Mu}, we again find that decreasing the field
               strength $B$ tends to push the closed-field line region
               outwards into the crust.}
\end{center}
\end{figure*}

Figure \ref{pol_Mu} suggests that the interior field structure of a
superconducting star can change dramatically as the field strength is
lowered, with the closed-field line region being `expelled' from the
core into the crust. Note that this is \emph{not} the Meissner effect!
To check that this result is not a peculiarity of
the functional choice $M_N(u)=\kappa u$, we plot a
sequence of equilibria for $M_N(u)=\kappa u^2$ in figure
\ref{pol_Muu}. From left to right we show `high' ($\langle
B^\textrm{cc}\rangle>H^\textrm{cc}_{c1}$), `medium' ($\langle B^\textrm{cc}\rangle\sim
H^\textrm{cc}_{c1}$) and `low' ($\langle B^\textrm{cc}\rangle<H^\textrm{cc}_{c1}$) 
field strengths, as in figure \ref{pol_Mu}. Although the interior
field distribution is different, the general trend for the closed
field lines to be pushed out into the crust remains.

\begin{figure}
\begin{center}
\begin{minipage}[c]{0.8\linewidth}
\includegraphics[width=\linewidth]{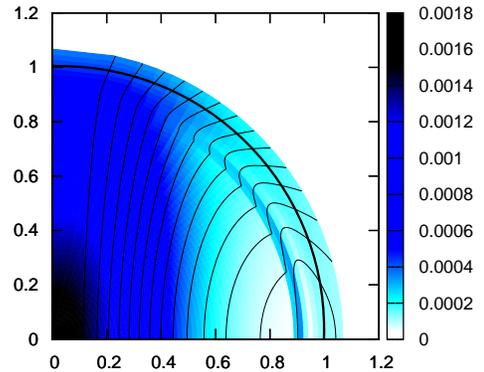}
\end{minipage}
\caption{\label{pol_strat}
               Poloidal field in a star with composition-gradient
               stratification in the core. This is a `low-field'
               model, with $\langle B\rangle<H_{c1}$ at the crust-core boundary. The
               neutron and proton polytropic indices are $N_\rmn=0.6$ and $N_\rmp=1.5$.}
\end{center}
\end{figure}

The sequences of equilibria shown so far are non-rotating and
unstratified, with $N_\rmn=N_\rmp=1$. For the last
two equilibrium models in this section, we study the effect of relaxing these
restrictions, both in the low-field case $\langle
B^\textrm{cc}\rangle<H^\textrm{cc}_{c1}$. Firstly we consider a star with
composition-gradient stratification in the core, matched to a
single-fluid unstratified crust as always. We take $N_\rmn=0.6$ and
$N_\rmp=1.5$, since an earlier study \citep{LAG} found that models with
$N_\rmn<N_\rmp$ mimic realistic equations of state more closely than
those with $N_\rmn\geq N_\rmp$. The resultant field configuration --- figure
\ref{pol_strat} --- is rather similar to that of the unstratified star
shown in the right-hand panel of figure \ref{pol_Mu}.

\begin{figure}
\begin{center}
\begin{minipage}[c]{0.8\linewidth}
\includegraphics[width=\linewidth]{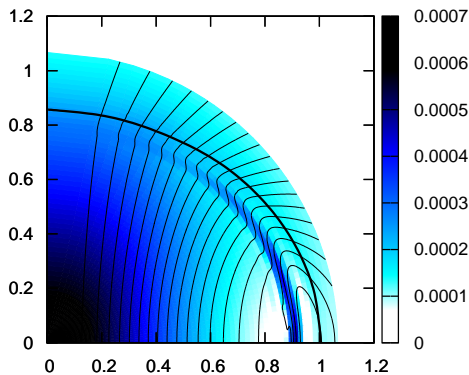}
\end{minipage}
\caption{\label{pol_rot}
               A rapidly-rotating unstratified star, again with a
               purely poloidal field and $\langle
               B\rangle<H_{c1}$ at the crust-core boundary. Rotation
               distorts the shape of the stellar surface and
               crust-core boundary, but otherwise the configuration is
               qualitatively similar to the other `low-field' models shown
               in this section.}
\end{center}
\end{figure}

Our last poloidal-field equilibrium is for a rapidly-rotating
unstratified star, with both neutron and proton fluids corotating
rigidly. Rotation in a superconductor induces a new contribution to
the magnetic field known as the London field --- but since this is
predicted to be around 1 G or less \citep{sauls_review,GAS} it is completely negligible for our
purposes. Our rotating model, whose field configuration is shown in figure
\ref{pol_rot}, is noticeably oblate, with the ratio of the polar to
equatorial surface radii $r_\textrm{pole}/r_{\textrm{eq}}=0.85$. For our canonical NS
parameters this corresponds to a rotation rate of 1000 Hz. The stellar
rotation also distorts the crust-core boundary away from a spherical
shape, and this is the main difference in appearance between this
model and the right-hand side model of figure \ref{pol_Mu}. The closed
field lines remain confined to the crust.

Finally, we calculate the magnetically-induced distortion within our
models. This serves two purposes: the qualitative scaling
provides a sanity check of our results and the quantitative results give us an
indication of the possible strength of gravitational radiation from a
magnetically-distorted neutron star \citep{bon_gour}. First we recall the definition
of the mass quadrupole moment tensor:
\beq
Q_{jk} = \int\rho x_j x_k \ \rmd V
\eeq
where $x_j,x_k$ are Cartesian coordinates. As a measure of the
distortion of the star's mass distribution, we now define the ellipticity
of the star through the components of the quadrupole moment
at the equator $Q_{xx}$ and pole $Q_{zz}$:
\beq
\epsilon=\frac{Q_{xx}-Q_{zz}}{Q_{xx}}.
\eeq

\begin{figure}
\begin{center}
\begin{minipage}[c]{\linewidth}
\psfrag{a}{$(a)$}
\psfrag{b}{$(b)$}
\psfrag{c}{$(c)$}
\psfrag{B_16}{$B_\textrm{pole}/10^{16}$ G}
\psfrag{ellip}{$\eps$}
\includegraphics[width=\linewidth]{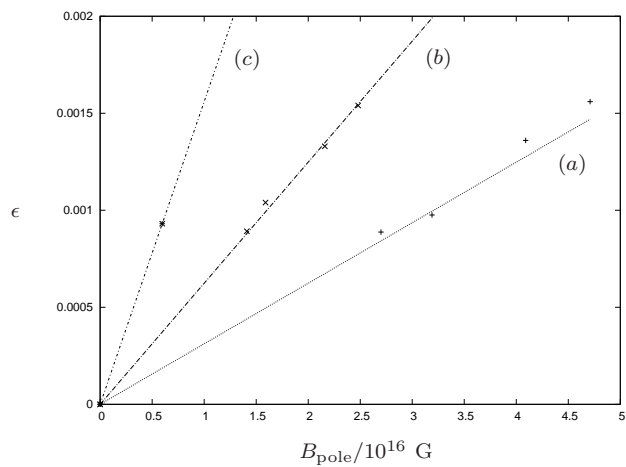}
\end{minipage}
\caption{\label{ellips}
              Scaling of magnetically-induced ellipticity $\eps$ with
              field strength at the polar cap $B_\textrm{pole}$. Lines
              $(a),(b)$ and $(c)$ are for models with a central
              critical field $H_{c1}(0)=1,2,5\times 10^{16}$ G
              respectively; these strong fields are needed to produce
              non-rotating stars with a sufficiently large distortion
              to be resolved on our numerical grid. Our results agree well with the
              expected scaling $\eps\sim H_{c1}B$.}
\end{center}
\end{figure}

We expect the ellipticity to scale with the magnetic energy,
i.e. $\eps\sim H_{c1}B$, as opposed to $\eps\sim B^2$ for normal
matter \citep{jones75,easson_peth}. In figure \ref{ellips} we plot the
ellipticity against field strength $B$ for a variety of models, all with
$M_N=\kappa u$ but different central critical fields, $H_{c1}(0)=1,2,5\times 10^{16}$
G. The field strengths are very high, as we need the distortion to be
sufficiently large to resolve on our numerical grid. Most of our
models are in the `high-field' regime, with $B>H_{c1}$ at the
crust-core boundary, but the data point for $H_{c1}(0)=5\times
10^{16}$ G represents a `medium-field' model. We use this point to fix
the gradient of line $(c)$, and plot lines $(a)$ and $(b)$ at $\frac{1}{5}$
and $\frac{2}{5}$ of this gradient. These are seen to agree well with the data
points for $H_{c1}(0)=1\times 10^{16}$ G and $H_{c1}(0)=2\times
10^{16}$ G, so we conclude that our results do indeed have the correct
ellipticity scaling.

Using figure \ref{ellips} and a similar plot for $M_N(u)=\kappa u^2$,
we find the following ellipticity relations for a neutron star with a
superconducting core:
\beq
\eps  = 3.1\times 10^{-8}\brac{\frac{B_\textrm{pole}}{10^{12}\ \textrm{G}}}
                                         \brac{\frac{H_{c1}(0)}{10^{16}\ \textrm{G}}}
\eeq
for $M_N(u)=\kappa u$, and 
\beq
\eps  = 4.4\times 10^{-8}\brac{\frac{B_\textrm{pole}}{10^{12}\ \textrm{G}}}
                                         \brac{\frac{H_{c1}(0)}{10^{16}\ \textrm{G}}}
\eeq
for $M_N(u)=\kappa u^2$. Note that our new boundary condition has
increased these values somewhat with respect to those reported in
\citet{sc_eqm_letter}, where the prefactors were $2.5\times 10^{-8}$
for $M_N(u)=\kappa u$ and $3.4\times 10^{-8}$ for $M_N(u)=\kappa
u^2$. An ellipticity often quoted for superconducting stars is the
estimate for purely toroidal fields given by \citet{cutler}, where the
ellipticity prefactor for $(B/10^{12}\textrm{ G})$ is $-1.6\times 10^{-9}$
(the negative sign reflects the prolate distortion which toroidal
fields induce). We see that our poloidal-field results give
distortions around 20 times bigger.

\subsection{Mixed poloidal-toroidal fields}

\begin{figure}
\begin{center}
\includegraphics[width=0.8\linewidth]{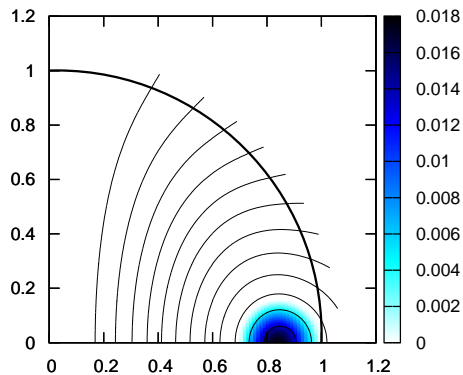}
\caption{\label{normal_mix}
               A mixed poloidal-toroidal magnetic field in a two-fluid
               core/single-fluid crust model, obeying normal MHD in both
               domains. We plot poloidal field lines, but this time
               the colour scale shows the magnitude of the toroidal
               field component only.}
\end{center}
\end{figure}

In the previous section we found that the ratio between the two fields
$B$ and $H_{c1}$ at the crust-core boundary seemed to be the most
important quantity dictating the structure of the poloidal field,
giving us three classes of equilibria: `low-field', 'medium-field' and
`high-field', depending on whether $\langle B^\textrm{cc}\rangle$ was
smaller than, comparable to, or greater than $H^\textrm{cc}_{c1}$. We now
turn to the general mixed-field case, where a toroidal component
is present as well as the poloidal one. Since we found that
neither stratification nor rotation have a significant effect on the
field structure, we consider only unstratified and non-rotating models
here.

As in the last section, we begin by presenting a typical field
configuration in a NS with normal matter in the core and crust, for
later comparison. Figure \ref{normal_mix} shows a mixed
poloidal-toroidal field, sometimes called a
`twisted-torus field' for its closed-field line geometry. The poloidal
field is present throughout the interior and extends outside the star,
whilst the toroidal component only exists in the small region of
closed field lines, as dictated by equation \eqref{fN_choice}. The maximum
value of the toroidal component is comparable with that of the
poloidal component, but it is globally weak, in the sense that it
occupies a small volume; it only contributes $3\%$ of the star's
magnetic energy.

\begin{figure*}
\begin{center}
\begin{minipage}[c]{\linewidth}
\includegraphics[width=\linewidth]{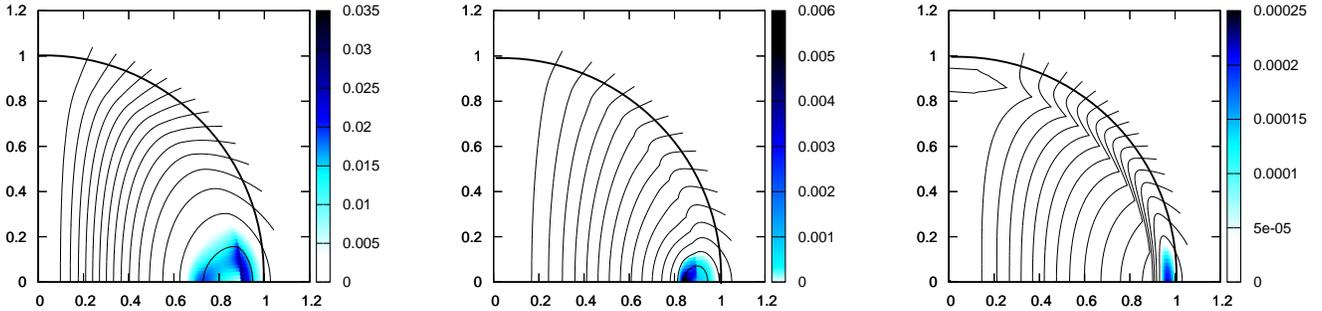}
\end{minipage}
\caption{\label{mixed}
               Mixed poloidal-toroidal field configurations in stars
               with a superconducting core and normal crust. We plot
               the poloidal field lines and use the colour scale to
               show the magnitude of the toroidal component only. As usual
               the plots, from left to right, show the `high',
               `medium' and `low' field cases (determined by the ratio
               $\langle B^\textrm{cc}\rangle/H^\textrm{cc}_{c1}$). The fate
               of the toroidal-field component is tied to that of the
               closed field lines, through equations \eqref{fsc-from-fN}
               and \eqref{fN_choice} --- and so it too is seen to
               be expelled from the core and confined to the crust as
               the field strength $B$ is decreased. In each case the
               toroidal component is locally strong but globally weak:
               its magnitude is comparable with that of the poloidal
               component in the same region, but its contribution to
               the magnetic energy is below $1\%$.}
\end{center}
\end{figure*}

In a superconducting star, the toroidal field must still be fitted
inside closed field lines (equation \eqref{fsc-from-fN}). Based on the
very small closed-field line regions seen in the last section, we may
therefore anticipate that any toroidal components will be confined to
a tiny volume of the star. This is borne out by figure
\ref{mixed}. The high-field model on the left-hand side is reminiscent of
the normal-matter case shown in figure \ref{normal_mix} --- although
whereas the toroidal field in the normal-matter star attains its
maximum at the centre of the closed-field line region, in the
superconducting star it vanishes there. In three dimensions, then, the
toroidal component forms a hollow tube, as also seen in
\citet{sc_eqm_letter}. This structure is in fact dictated by the form
of the toroidal-field flux function; one can rearrange equation
\eqref{f_supercon}, using
$\Pi=B/\rhop=\sqrt{B_\textrm{pol}^2+B_\phi^2}/\rhop$, to show that
\beq
B_\phi = \frac{B_\textrm{pol}f(u)}{\sqrt{\rhop^2\varpi^2-f^2(u)}}
\eeq 
--- when $B_\textrm{pol}$ vanishes, $B_\phi$ must too.

Moving to the middle panel of figure \ref{mixed}, for a `medium-field' model, we see that the
toroidal component has been pushed outwards, and in the right-hand
panel --- a `low-field' star --- the toroidal component exists only in the
normal crust. In all three plots the toroidal field is locally comparable with
the poloidal component, but globally extremely weak: from left to
right, the percentages of magnetic energy in the toroidal component are
$0.11, 0.0025$ and $2.2\times 10^{-6}$ respectively.

\section{Discussion}

The main conclusion of this work is that field configurations inside a
superconducting neutron star may vary dramatically depending on the field
strength, in contrast to normal-matter models where the field geometry
is essentially independent of its strength. The reason is that, unlike
normal matter, type-II superconductors have a characteristic field strength $H_{c1}\sim
10^{15}$ G, which is density-dependent and enters the
governing equations of the system. Our results indicate that the
crucial quantity dictating the equilibrium configuration is the ratio of $B$ to $H_{c1}$ at the crust-core
boundary.

\begin{table*}
\begin{center}
\caption{\label{infer_B}
               Expected features of neutron star internal magnetic
               fields, for different polar-cap field strengths, based
               on our models. These results are for $M_N(u)=\kappa u$,
               but the case $M_N(u)=\kappa u^2$ is qualitatively similar.}
\begin{tabular}{cccc}
\hline
polar-cap field & $H^\textrm{cc}_{c1}$ vs $B^\textrm{cc}$ & $\bar{B}/B_\textrm{pole}$ & closed-field-line/toroidal-field region \\
\hline
$B_\textrm{pole}\lesssim 5\times 10^{13}$ G   
        & `low': $\langle B^\textrm{cc}\rangle\lesssim 0.4H^\textrm{cc}_{c1}$ & $6-7$ & confined to crust \\
$2\times 10^{14}\textrm{G}\lesssim B_\textrm{pole}\lesssim 2\times 10^{15}$ G
        & `medium': $0.4H^\textrm{cc}_{c1}\lesssim\langle B^\textrm{cc}\rangle\lesssim H^\textrm{cc}_{c1}$ & $1.2-1.8$ & in crust and core, centred in crust \\
$2\times 10^{15}\textrm{G}\lesssim B_\textrm{pole}\lesssim 6\times 10^{15}$ G  
        & `high': $\langle B^\textrm{cc}\rangle\gtrsim H^\textrm{cc}_{c1},\bar{B}<H_{c2}$ & $1.7$ & in crust and core, centred in core \\
$B_\textrm{pole}\gtrsim 5\times 10^{15}$ G  
        & normal matter: $\bar{B}>H_{c2}$ & $1.9$ & in crust and core, centred in core \\
\hline
\end{tabular}\\
\end{center}
\end{table*}

Within the context of our model, we report in table \ref{infer_B}
the key features we would expect a neutron star's interior field to have,
based on its polar-cap field. This serves as a summary of the main
results of this paper. For polar-cap fields below $\sim 5\times
10^{13}$ G we expect discontinuities at the crust-core boundary and
any toroidal component to be confined to the crust. The
volume-averaged field for these `low'-field models may be a factor of
7 stronger than than the observed polar-cap
value. Our scheme has not converged to any models in the range
$6\times 10^{13}\textrm{G}\lesssim B_\textrm{pole}\lesssim 2\times 10^{14}$ G,
presumably because of numerical limitations. Our `medium'-field
models start at around $2\times 10^{14}$ G and span an order of
magnitude, covering typical magnetar values; for these configurations $B_\textrm{pole}$
is typically about half of the average interior strength and the toroidal
component begins to penetrate the core. In the `high'-field range, up
to $\sim 6\times 10^{15}$ G, the toroidal component pervades more of
the interior and the configuration begins to resemble the smooth
solutions from stars with normal matter in the core and crust.

Superconductivity is destroyed for magnetic fields above the upper
critical field $H_{c2}$, at which point one returns to the regime of
normal MHD \citep{GAS}. This critical field is density-dependent, but
for simplicity we will --- rather arbitrarily --- assume the
destruction of superconductivity happens when the volume-averaged
field $\bar{B}>10^{16}$ G. Using our earlier code for a star with a
normal-matter core \citep{LAG}, we would expect this interior
field to be attained for a polar-cap field of $B_\textrm{pole}\sim 5\times
10^{15}$ G. Now using our superconducting models for an
independent estimate of this field strength, we arrive at
a reassuring similar value of $B_\textrm{pole}\sim 6\times 10^{15}$ G.

Our naive estimates suggest that no known magnetar has a field quite high enough
to destroy superconductivity in the interior; the highest observed polar-cap
field is $2\times 10^{15}$ G, for SGR 1806-20. Many of them
\emph{are}, however, in a range around $B_\textrm{pole}\sim 10^{14}$ G,
corresponding roughly to the strength separating the qualitatively
different `low' and `medium'-field configurations in our work. As a
magnetar's field decays, our models indicate that it will need to
rearrange substantially to adapt to its new equilibrium, with a
surface field strength considerably lower than the interior
average. This \skl{predicted rearrangement} may play a role in the flaring and
glitch activity observed for many magnetars.

In all our models the toroidal field component can be locally important
(comparable in magnitude to the poloidal one), but is always globally
insignificant. Accordingly, the toroidal component plays no role in
the size of magnetically-induced distortions, and we expect it to be
irrelevant for oscillations of our model neutron stars too. It may be
relevant for a neutron star's stability, however, as magnetic instabilities are
typically local in nature. Recent work for normal-matter stars
\citep{ciolfi_rezz} has found solutions with strong toroidal
components by careful choices of the magnetic free functions entering
the governing equations, but it seems less likely this would work in
the context of a superconducting star, where the physics at the
crust-core boundary has the strongest effect on the resultant equilibria.

The results in this paper suggest intriguing differences between the
interior fields of neutron stars in the typical range for pulsars,
$B\lesssim 5\times 10^{13}$ G, and those in the magnetar range, $B\gtrsim
10^{14}$ G. At the same time, we should be cautious that our modelling of neutron star
superconductivity here is quite primitive. Essentially, we have only managed to
account for the basic features of a model \citep{baym_pp} which is now
more than forty years old! The most obvious \skl{pieces of} physics missing from our
models are entrainment --- which results in an effective magnetic force
on the neutrons in addition to the one on the protons --- and
interactions between neighbouring fluxtubes, expected for the
range $H_{c1}<B<H_{c2}$ when fluxtubes become densely packed
within the core. 

In addition, neutron stars are likely to have an inner
region where the superconductivity is of type I \citep{jones06,GAS}, or could even
be type-I throughout the core \citep{buckley_mz}. At present there
is no set of macroscopic smooth-averaged equations for this case,
as would be needed to find equilibria in our approach. There may also be
exotic physics at the transition between type-II and type-I
superconductivity \citep{alford_good}. Finally, if the density is high enough there could
be an inner core of exotic matter, like hyperons or quarks \citep{alcock}. If we
want a better understanding of magnetic-field physics in neutron
stars, these difficult but interesting issues must be confronted.

\section*{Acknowledgements}

I am pleased to thank Nils Andersson for his comments on the results
presented here. This work is supported by the German Science
Foundation (DFG) via SFB/TR7.

\appendix

\section{Derivation of the `Grad-Shafranov' equation for a superconductor}
\label{supercon_GS_deriv}

In section \ref{general_eqs} we established the basic results of axisymmetric MHD, which are
independent of the form of the magnetic force. We now specialise to the
fluxtube tension force discussed in section \ref{supercon_summary}:
\beq \label{orig_fmag}
\fmag = -\frac{1}{4\pi}\Bigg[
                  \bB\times(\curl{\bf H}_{c1})
                   +\rhop\nabla\brac{B\pd{H_{c1}}{\rho_\rmp}}
                                  \Bigg],
\eeq
where ${\bf H}_{c1}=H_{c1}\hat\bB$ is the critical field, pointing in
the same direction as $\bB$. To our order of working
$H_{c1}=h_c\rho_\rmp/\ve_*$, where $h_c$ and
$\ve_*$ are constants \citep{GAS}. We will
assume zero entrainment (meaning $\ve_*=1$), otherwise there will be
a magnetic force acting on the neutrons too. Equation \eqref{orig_fmag} now becomes
\beq \label{early_fmag}
-\frac{4\pi}{h_c}\fmag 
   = \rho_\rmp\nabla B 
      + \bB\times\brac{ \rho_\rmp\curl\hat\bB
                                    + \nabla\rho_\rmp\times\hat\bB }.
\eeq
At this point it is convenient to define a `unit current'
$\uj\equiv\curl\hat\bB$; as in the single-fluid case \citep{LJ09} it may be shown that
\beq
\uj = \frac{1}{\pom}\nabla(\pom\hat{B}_\phi)\times\be_\phi
            + \hat{\j}_\phi\be_\phi.
\eeq
Decomposing the bracketed terms from the right-hand side of equation
\eqref{early_fmag} into poloidal and toroidal components gives, after
some rearrangement,
\begin{align}
\rho_\rmp\curl\hat\bB + \nabla\rho_\rmp\times\hat\bB =& 
\frac{1}{\pom}\nabla(\rho_\rmp\pom\hat{B}_\phi)\times\be_\phi\nn\\
  & + \brac{\rho_\rmp\hat\j_\phi
               - \frac{\nabla\rho_\rmp\cdot\nabla u}{\pom B} }\be_\phi
\end{align}
so the magnetic force becomes
\beqa
-\frac{4\pi}{h_c}\fmag 
   =& \rho_\rmp\nabla B
             +\frac{1}{\pom}\bB
                 \times\brac{\nabla(\rho_\rmp\pom\hat{B}_\phi)\times\be_\phi}\nn\\
    &     +\brac{\rho_\rmp\hat\j_\phi
                - \frac{\nabla\rho_\rmp\cdot\nabla u}{\pom B} }\bB\times\be_\phi.
\label{fmag_interim}
\end{align}
Noting that
\beq
\bB\times\be_\phi=\bB_\textrm{pol}\times\be_\phi=-\frac{\nabla u}{\pom},
\eeq
and using standard vector identities, we may rearrange
\eqref{fmag_interim} into the form:
\beqa
-\frac{4\pi}{h_c}\fmag 
   =& \rho_\rmp\nabla B
           +\frac{1}{\pom^2}\nabla u\times
                      \nabla(\rho_\rmp\pom\hat{B}_\phi)
           +\frac{B_\phi}{\pom}\nabla(\rho_\rmp\pom\hat{B}_\phi)\nn\\
     &  +\brac{\frac{\nabla\rho_\rmp\cdot\nabla u}{\pom B}
                  -\rho_\rmp\hat\j_\phi}\frac{\nabla u}{\pom}.
\end{align}
In axisymmetry, $\fmag$ has no toroidal component. Examining the above expression, we see all terms are poloidal
except the following one, which must therefore be zero:
\beq \label{parallelnablas}
\frac{1}{\pom^2}\nabla u\times\nabla(\rho_\rmp\pom\hat{B}_\phi)=0.
\eeq
Hence we can remove this term from the magnetic force to arrive at our
`interim' result for $\fmag$:
\beqa
-\frac{4\pi}{h_c}\fmag 
   =& \rho_\rmp\nabla B
            +\frac{B_\phi}{\pom}\nabla(\rho_\rmp\pom\hat{B}_\phi)\nn\\
     &    +\brac{\frac{\nabla\rho_\rmp\cdot\nabla u}{\pom B}
                     -\rho_\rmp\hat\j_\phi}\frac{\nabla u}{\pom}.
 \label{fmag_general}
\end{align}
We now have the same dichotomy as for the normal-MHD case:
satisfying \eqref{parallelnablas} leads to a mixed-field case, or a purely-toroidal case. For the former,
we require that $\nabla u$ and $\nabla(\rho_\rmp\pom\hat{B}_\phi)$
be parallel, which leads to
\beq \label{f_sc}
\rho_\rmp\pom\hat{B}_\phi = f(u)
\eeq
for some function $f$. In the special case $f(u)=0$ the field is
purely poloidal. Alternatively, we can satisfy \eqref{parallelnablas}
by taking $\nabla u=0$, so that our field is purely toroidal. We begin
with this latter, simpler case.

\subsection{Toroidal fields}

The purely-toroidal field case is no more involved than the
corresponding derivation for normal matter. It has been studied in
earlier papers \citep{akgun_wass,LAG}, but we include the results for
completeness. From \eqref{fmag_general}, we have $\nabla u=0$ and so the
expression for the magnetic force reduces to
\beq \label{fmag_tor}
\fmag = -\frac{h_c}{4\pi} 
                       \frac{1}{\pom} \nabla(\pom\rho_\rmp B_\phi).
\eeq
Now since $\curl(\fmag/\rho_\rmp)=0$ we have
\beq
0 = \nabla(\pom\rho_\rmp)\times\nabla(\pom\rho_\rmp B_\phi)
\eeq
and hence the arguments of the two gradient operators must be related by some
function $\eta$:
\beq
\eta(\pom\rho_\rmp) = \pom\rho_\rmp B_\phi.
\eeq
Putting this into \eqref{fmag_tor}, using $\fmag=\rho_\rmp\nabla M$ and
defining $\zeta\equiv\pom\rho_\rmp$ we get
\beq \label{nabM_tor}
\nabla M = -\frac{h_c}{4\pi}\frac{1}{\zeta}\nabla\eta(\zeta)
               = -\frac{h_c}{4\pi}\frac{1}{\zeta}\td{\eta}{\zeta}\nabla\zeta
\eeq
where the corresponding magnetic field is
\beq
\bB=B_\phi\be_\phi=\frac{\eta(\zeta)}{\zeta}\be_\phi.
\eeq
Equation \eqref{nabM_tor} gives the magnetic force in a form that
allows for direct integration of the proton Euler equation and
no additional equation for the magnetic field is needed.

\subsection{Mixed poloidal-toroidal fields}

We return to the general equation for the magnetic force
\eqref{fmag_general}. The $B_\phi$ terms in this expression may be
replaced using the relation \eqref{f_sc} and the left-hand side may be
rewritten using $\fmag=\rho_\rmp\nabla M$. Now applying the chain rule
to the resulting $\nabla f(u)$ term and rearranging, we have
\beq \label{nablas}
-\frac{4\pi}{h_c}\nabla M-\nabla B
  = \brac{ \frac{Bf}{\rhop^2\varpi}\td{f}{u}
                + \frac{\nabla\rho_\rmp\cdot\nabla u}{\rho_\rmp\pom B}
                - \hat\j_\phi } \frac{\nabla u}{\pom}.
\eeq
At this point in the normal-matter derivation one dots $\bB$ into this
equation, showing that $\bB\cdot\nabla M=0$; together with
$\bB\cdot\nabla u=0$, this gives the important result $M=M(u)$. For
superconducting matter this is no longer true. Instead, let us
define a new function
\beq \label{y_defn}
y = \frac{4\pi M}{h_c}+B.
\eeq
Now, dotting $\bB$ into equation \eqref{nablas} we have
\beq
-\bB\cdot\nabla y
  = \frac{1}{\pom}
       \brac{ \frac{Bf}{\rhop^2\varpi}\td{f}{u}
              + \frac{\nabla\rho_\rmp\cdot\nabla u}{\rho_\rmp\pom B}
              - \hat\j_\phi } \bB\cdot\nabla u = 0
\eeq
--- we know that $\bB\cdot\nabla u=0$, and so $\bB\cdot\nabla y=0$
too. But this in turn means that $\nabla u$ and $\nabla y$ are
parallel, so:
\beq
y=y(u)
\eeq
and hence
\beq
-\td{y}{u}\nabla u
  = \brac{ \frac{Bf}{\rhop^2\varpi}\td{f}{u}
                + \frac{\nabla\rho_\rmp\cdot\nabla u}{\rho_\rmp\pom B}
                - \hat\j_\phi } \frac{\nabla u}{\pom}.
\eeq
For $\nabla u\neq 0$ we then get a relation between
the functions $y$ and $f$:
\beq \label{y_eq}
y'(u) = -\frac{B}{\varpi^2\rhop^2}f(u)f'(u)
            -\frac{\nabla\rhop\cdot\nabla u}{\varpi^2\rhop B}
            +\frac{\hat\j_\phi}{\varpi}.
\eeq
Let us now get rid of $\hat\j_\phi$ in favour of an expression in terms
of $u$:
\beq
\hat\j_\phi
  =[\curl\hat\bB]_\phi
  = -\frac{1}{\varpi B}\brac{\Delta_* u
                                           - \frac{\nabla B\cdot\nabla u}{B}},
\eeq
using the same $\Delta_*$ operator as for the Grad-Shafranov equation
\eqref{grad_shaf}. As for normal MHD, this step gives $\hat\j_\phi$ in
terms of a differential operator acting on $u$ (albeit a more
complicated one). Now using this to replace $\hat\j_\phi$ in equation
\eqref{y_eq}, we have
\beq
\Delta_* u -\frac{\nabla B\cdot\nabla u}{B}
            +\frac{\nabla\rhop\cdot\nabla u}{\rhop}
  = -\frac{B^2}{\rhop^2}ff'-\varpi^2 B y'.      
\eeq
Since the quantity $B/\rho_\rmp$ appears frequently in the above
expression, let us call this $\Pi$. One final rearrangement of the
magnetic force expression then gives our final result, a
`Grad-Shafranov' equation for type-II superconductors:
\beq \label{supercon_GS_appendix}
\Delta_* u -\frac{\nabla\Pi\cdot\nabla u}{\Pi}
  = -\varpi^2\rhop\Pi y' - \Pi^2 ff'.
\eeq
One could now replace the $\Pi$ terms by quantities involving $u$, since
\beq \label{Pi-u}
\Pi \equiv \frac{B}{\rho_\rmp}
       = \frac{|\nabla u|}{\sqrt{\varpi^2\rhop^2-f^2}},
\eeq
but the result is forbidding and not obviously useful.

\section{Numerical method}
\label{appendix_numerics}

\subsection{The differential operator in equation \eqref{supercon_GS_appendix}}

As written in equation \eqref{supercon_GS_appendix}, our problem
features a very unfamiliar differential operator
$(\Delta_*-\frac{1}{\Pi}\nabla\Pi\cdot\nabla)$ acting on the
streamfunction, and we would have to find a Green's function to invert
it. Even the normal-matter Grad-Shafranov equation \eqref{grad_shaf}, however, is
unconventional in having $u$ dependence on the left and right-hand
sides. It seems equally legitimate to move the $\nabla\Pi\cdot\nabla
u/\Pi$ term to the right-hand side, especially since we will use an
iterative scheme which should gradually update both sides of the
equation and approach a consistent solution. Whether or not this
approach is acceptable can be decided based on if the scheme
converges, and if the resultant configuration satisfies the
equilibrium form of the virial theorem, discussed later in this section.

Accordingly, we now proceed in a similar manner to that previously
used to solve the Grad-Shafranov equation \citep{tomi_eri,LJ09},
converting the magnetic equation \eqref{supercon_GS_appendix} into a
Poisson-like equation, which may then be solved using familiar
Green's-function methods. For this we use the following relation between the
$\Delta_*$ and Laplacian operators:
\beq \label{GStoLap}
\Delta_* u=\frac{\varpi}{\sin\phi}\nabla^2\brac{\frac{u\sin\phi}{\varpi}}.
\eeq
Equation \eqref{supercon_GS_appendix} then takes the form:
\begin{align}
\nabla^2\brac{\frac{u\sin\phi}{\varpi}}
 = -\left[-\frac{\nabla\Pi\cdot\nabla u}{\varpi\Pi}  \right. & +\varpi\rhop\Pi y'(u) \nn\\
                            + & \left. \frac{\Pi^2}{\varpi}f(u)f'(u)\right]\sin\phi.
\end{align}
In integral form this is:
\begin{align}
u = \frac{\varpi}{4\pi\sin\phi}\int
         \left[-\frac{\nabla\tilde\Pi\cdot\nabla\tilde{u}}{\tilde\varpi\tilde\Pi}\right. &
                  +\tilde\varpi\tilde\rhop\tilde\Pi y'(\tilde{u}) \nn\\
                            +&\left. \frac{\tilde\Pi^2}{\tilde\varpi}f(\tilde{u})f'(\tilde{u})\right]
               \frac{\sin\tilde\phi}{|\br-\tilde\br|}\ \rmd\tilde\br,
               \label{supercon_uint}
\end{align}
denoting dummy variables with tildes.

\subsection{Iterative scheme}

As in \citet{LAG}, our numerical method is based on the Hachisu
self-consistent field method \citep{hachisu}, an iterative scheme originally used for rotating
unmagnetised stars. The method finds
solutions to the equilibrium equations in integral form, by
progressively updating the density distribution to satisfy some
user-specified surface distortion. For stars with magnetic fields and
normal matter, it is quite easily generalised by including the
integral form of the Grad-Shafranov equation in the scheme
\citep{tomi_eri,LAG}, and is robust. The extra nonlinearities of the
new magnetic equation \eqref{supercon_uint} cause the method
difficulties, however --- with direct implementation resulting in a non-convergent,
unstable scheme. We have found these can be cured by
`normalising' the magnetic integral by the maximum value attained by
$\Pi$ (dividing by $\Pi_\textrm{max}$ when solving the integral and
multiplying again afterwards) and by using underrelaxation. 

The underrelaxation step works by updating the streamfunction only
partially at each iterative step. If $u_\textrm{new}^*$ is the result of integrating
\eqref{supercon_uint}, then \emph{fully} updating the streamfunction
would entail discarding the form $u_\textrm{old}$ from the last iterative
step and setting $u_\textrm{new}=u_\textrm{new}^*$ for the new streamfunction;
instead we update by using
\beq \label{underrelax}
u_\textrm{new} = (1-\omega)u_\textrm{old} + \omega u_\textrm{new}^* .
\eeq
Clearly $\omega=1$ gives us the conventional full-update case. When
solving \eqref{supercon_uint}, we have found that the 
most successful value of the underrelaxation parameter $\omega$
varies with the grid resolution and the input parameters, but is typically
in the range $0.01-0.2$.

The complete iterative scheme, which converges successfully, takes the form:\\
0. For the initial conditions, start with simple trial guesses for
$\rho_\rmn,\rho_\rmp$ and $u$;\\
1. Calculate the gravitational potential $\Phi$ from the $\rho_\rmn$ and
$\rho_\rmp$ distributions and Poisson's equation \eqref{Poisson};\\
2. Calculate $\Pi$ from $u$ using \eqref{Pi-u};\\
3. Calculate the new magnetic streamfunction $u$ from its
value at the previous iteration $u_\textrm{old}$, using the magnetic
Poisson equation \eqref{supercon_uint} with $u_\textrm{old}$, $\Pi$ and $\rho_\rmp$ in the
integrand. Divide the integrand by $\Pi_\textrm{max}$ before evaluation,
multiply again afterwards and employ the underrelaxation step \eqref{underrelax};\\
4. Evaluate the proton-fluid Euler at the equatorial
and polar surfaces, using boundary conditions on $\mup$; this gives two
equations which fix $\Omega^2$ and then the integration constant for
the proton Euler equation;\\
5. We are now able to use the proton Euler to find $\mup$
throughout the star;\\
6. Evaluate the difference-Euler at the equatorial
neutron-fluid surface to find the integration constant for the
difference-Euler;\\
7. Now use the difference-Euler to find $\mun$ throughout the star;\\
8. Calculate the new density distributions from the chemical
potentials, as in \citet{LAG};\\
9. Return to step 1 using the new $\rho_\rmn,\rho_\rmp$ and $u$;
repeat procedure until satisfactory convergence is achieved,
i.e. until the fractional changes in quantities at consecutive iterative 
steps is less than some small tolerance (usually $10^{-4}$).\\

The input parameters for any equilibrium configuration are the surface
distortion $r_\textrm{pole}/r_{\textrm{eq}}$, the proton and neutron polytropic
indices $N_\rmp,N_\rmn$ and the constants $\kappa$ and $a$, related to the
strengths of the poloidal and toroidal field components. The iterative
process described here typically takes $100-1000$ steps. 
By comparison, the code for normal matter using in \citet{LAG}
usually converged within ten. We ensure the crust-core boundary is
well resolved by having a larger number of radial grid points than
angular ones, usually $480\times 180$.

\subsection{Tests of the numerical scheme}

To check that our resulting solutions are truly equilibria, we want to
use the virial theorem, as in the normal-MHD case.  The virial theorem
states that a certain combination of energies gives the acceleration
of the system (the second time-derivative of the moment of inertia
$I$):
\beq
2T+\clE_{mag}+W+3(\gamma_\rmn-1)U_\rmn+3(\gamma_\rmp-1)U_\rmp
 = \frac{1}{2}\td{^2 I}{t^2},
\eeq
for our double-polytrope EOS. For an equilibrium we want $\ddot{I}=0$,
up to the accuracy of our numerical scheme.

\begin{figure}
\begin{center}
\begin{minipage}[c]{0.9\linewidth}
\psfrag{virial test}{virial test}
\psfrag{radial grid points}{radial grid points}
\includegraphics[width=\linewidth]{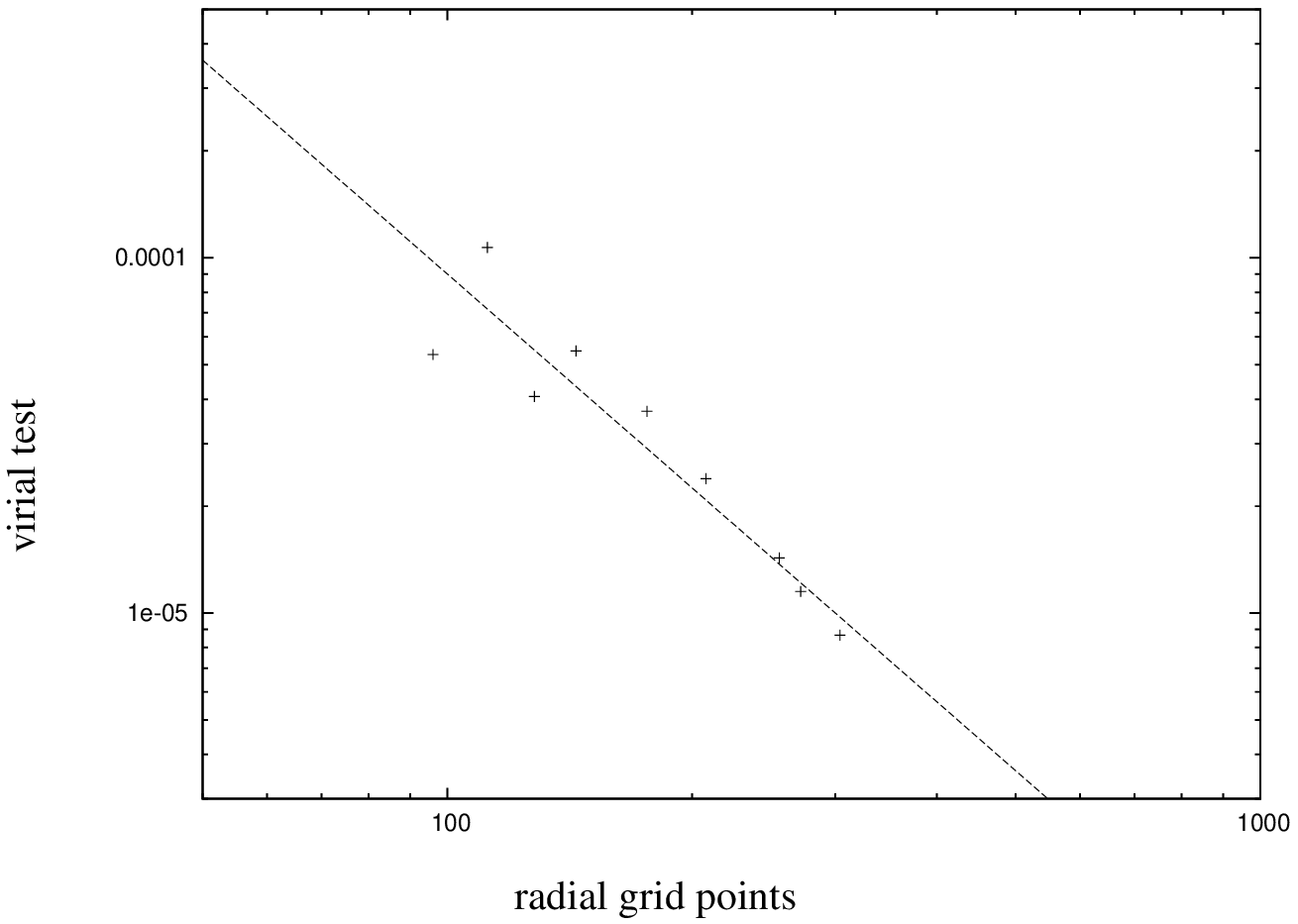}
\end{minipage}
\caption{\label{convergence}
               The virial test measures the deviation of our results
               from the expected value of zero for an equilibrium, and
               therefore quantifies our numerical error. We plot
               this for the same poloidal-field equilibrium at 
               different grid resolutions. The line shows the expected 
               scaling for second-order convergence; our results agree well with
               this.}
\end{center}
\end{figure}

Our results should approach the continuum solution, where
$\ddot{I}=0$, at infinite resolution. We can check this by using the
following quantity as a diagnostic:
\beq
\textrm{virial test}\equiv
\frac{2T+\clE_{mag}+W+3(\gamma_\rmn-1)U_\rmn+3(\gamma_\rmp-1)U_\rmp}{|W|}.
\eeq
This is plotted in figure \ref{convergence}, for different radial grid
resolutions. The plot shows that the results converge at
second order, the intended order of the code. In the main part of the paper we
also perform a more physical sanity check: checking that the
magnetically-induced distortion to the star has the correct scaling of
$\epsilon\sim H_{c1}B$. Figure \ref{ellips} shows that our results
agree convincingly with this relation.

\label{lastpage}

\end{document}